\documentclass[12pt]{article}
\usepackage{graphicx}
\usepackage{hyperref}

\def \bea{\begin{eqnarray}}
\def \beq{\begin{equation}}
\def \eea{\end{eqnarray}}
\def \eeq{\end{equation}}

\begin{document}
\rightline{EFI 12-5}
\rightline{arXiv:1205.1529}
\rightline{May 2012}

\bigskip
\centerline{\bf NON-FACTORIZABLE EFFECTS IN TOP QUARK PRODUCTION}
\bigskip

\centerline{Jonathan L. Rosner}
\centerline{\it Enrico Fermi Institute and Department of Physics}
\centerline{\it University of Chicago, 5620 S. Ellis Avenue, Chicago, IL 60637}
\bigskip

\begin{quote}
The production of top-antitop pairs at the Fermilab Tevatron shows a
forward-backward asymmetry in which the top quark tends to follow the proton
direction, while the antitop tends to follow the antiproton direction.  The
effect grows with increasing effective mass $m_{t \bar t}$ of the top-antitop
pair, and with increasing rapidity difference between the top and antitop.
The observed effect is about three times as large as predicted by
next-to-leading-order QCD, but with the same sign.  An estimate of
non-factorizable effects based on a QCD string picture finds they are
negligible, but that small distortions of the $m_t$ spectrum are possible.
Tests for such effects, both at and above the level of this estimate,
are suggested.
\end{quote}

\leftline{PACS numbers: 14.65.Ha, 12.38.Aw, 12.38.Qk, 12.39.St}
\bigskip

\section{INTRODUCTION}
The top quark has been shown to be produced at the Fermilab Tevatron with a
noticeable forward-backward asymmetry \cite{CDFAFB,Miet,D0AFB} in excess of
QCD predictions \cite{AFBth}.  This has spawned a number of possible
interpretations in terms of new physics \cite{Cao:2010zb}.

In the present paper we explore a different possibility.  According to the
factorization hypothesis, the subprocess $q \bar q \to t \bar t$ is assumed
to take place in isolation from the remaining constituents of the incident
proton and antiproton.  We investigate using a QCD string picture whether
the tendency for the top quark to follow the direction of the proton, and for
the antitop to follow the antiproton, may be due to nonperturbative QCD effects
violating factorization.  We find an effect much too small to account for the
observed asymmetry, but note the possibility of small distortions of the $m_t$
spectrum.  Such distortions are less likely at the CERN Large Hadron Collider
(LHC), where the dominant top pair subprocess is $g g \to t \bar t$.  Other
tests are suggested for non-factorizable effects.

We begin by reviewing the data and QCD predictions (Sec.\ II).  We then
examine in Sec.\ III a source of non-factorizable effects due to the
formation of QCD strings between the produced top or antitop and remaining
$p$ or $\bar p$ constituents.  The effects are found to be very small as a
result of the rapid top quark decay.  Qualitative
tests for other manifestations of the breakdown of factorization in hadronic
$t \bar t$ production include distortions in top reconstruction (Sec.\ IV),
enhanced multiparticle production between the $t$ and $p$ remnants or the
$\bar t$ and $\bar p$ remnants (Sec.\ V) and a forward-backward or charge
asymmetry in baryon/antibaryon production (Sec.\ VI). We conclude in Sec.\ VII.

\section{DATA AND QCD PREDICTION}

A top or antitop quark produced by $p \bar p$ reactions at the Tevatron
may be characterized by its rapidity $y \equiv (1/2)\ln[(E+P_z)/(E-p_z)]$,
where $E$ is its energy and $p_z$ the projection of its momentum along
the beam axis.  One may define $\Delta y \equiv y_t - y_{\bar t}$;
this variable is invariant under boosts along the $z$ axis and has the
same sign as $\cos \theta^*$, where $\theta^*$ is the angle the top
quark makes with the $+z$ direction (taken to be that of the proton)
in the $t \bar t$ center of mass system (c.m.s.) The forward-backward
asymmetry in top production at the parton level then may be defined as
\beq
A_{FB} \equiv \frac{N_{\Delta y > 0} - N_{\Delta y < 0}}
              {N_{\Delta y > 0} + N_{\Delta y < 0}}~.
\eeq
For the Tevatron running at $\sqrt{s} = 1.96$ GeV, the theoretical prediction
at non-leading QCD order and including electroweak corrections \cite{AFBth}
is $A_{FB} = 6.6\%$.  The most recent experimental data are:
\beq
A_{FB} = \left\{ \begin{array}{c} (16.2 \pm 4.7)\%~\cite{Miet} \cr
                 (19.6 \pm 6.5)\%~\cite{D0AFB} \end{array} \right.
\eeq
A linear growth of the asymmetry as a function of $\Delta y$ or $M_{t \bar t}$
is observed.  This behavior is also predicted by theory, but the slopes of
the effect are observed to be about three times the prediction.

At the LHC, the incident symmetric $pp$ collisions imply no forward-backward
asymmetry, but a charge asymmetry $A_C$ between $t$ and $\bar t$ is still
possible at next-to-leading order (NLO) in QCD, where
\beq
A_C \equiv \frac{N_{\Delta |y| > 0} - N_{\Delta |y| < 0}}
              {N_{\Delta |y| > 0} + N_{\Delta |y| < 0}}~,
\eeq
where $\Delta |y| \equiv |y_t| - |y_{\bar t}$.  The observed values are
$A_C = -0.018 \pm 0.028 \pm 0.023$ (ATLAS \cite{ATLAS:2012an}), 
$A_C = -0.013 \pm 0.028 ^{+0.029}_{-0.031}$ (CMS \cite{Chatrchyan:2011hk}), and
$A_C = 0.004 \pm 0.010 \pm 0.012$ (CMS \cite{CMSTOP11030}).
The theoretical prediction \cite{Kuhn:2011ri} is $A_C = 0.0115 \pm 0.0006$
(ATLAS \cite{ATLAS:2012an} quotes a slightly different value of $0.006 \pm
0.002$.)

\section{STRING-FRAGMENTATION PICTURE}

\begin{figure}
\begin{center}
\includegraphics[width=0.6\textwidth]{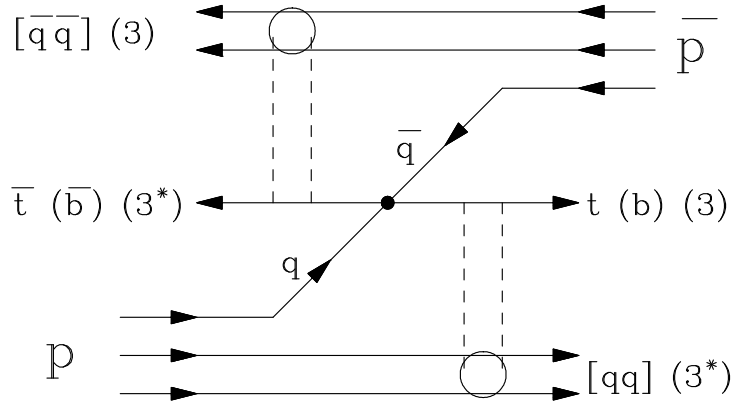}
\end{center}
\caption{Formation of QCD strings (dashed lines) between $t$ or $b$ (3 of
SU(3)$_{\rm C}$) and proton remnant ($3^*$ of SU(3)$_{\rm C}$), and between
$\bar t$ or $\bar b$ ($3^*$) and antiproton remnant (3).
\label{fig:drag}}
\end{figure}

The mirroring of the qualitative behavior of $A_{FB}$ for top pair production
in $p \bar p$ collisions by that of perturbative QCD suggests a possible
origin in nonperturbative effects.  One such mechanism is illustrated in
Fig.\ \ref{fig:drag}.

At the Tevatron, the dominant subprocess in top-antitop production is 
$q \bar q \to t \bar t$, with quarks $q$ dominantly coming from the proton
and $\bar q$ from the antiproton.  A proton giving up a quark is left as
a color antitriplet remnant, while an antiproton giving up an antiquark is
left as a color triplet remnant.

A top quark is expected to decay via $t \to b W^+$ with a lifetime of about
$5 \times 10^{-25}$ s \cite{LissPDG}.  The top or its daughter $b$ can
interact during the initial stages of the fragmentation process with the
remnant of either the proton or the antiproton; either interaction is
expected to affect its angular distribution.  Within the factorization
hypothesis, this interaction is expected to be small and to be successfully
described by perturbative QCD corrections at next-to-leading order.

A one-gluon exchange potential has a definite form between two color triplets
in a 6 or $3^*$ representation of color SU(3), or a triplet and antitriplet in
an octet or singlet \cite{Nambu,Lipkin73,StX}:
\bea
V_{33;3^*}&=&-\frac{2}{3} \frac{\alpha_S}{r}~;~~
V_{33;6}   =  \frac{1}{3} \frac{\alpha_S}{r}~;~~ \nonumber \cr
V_{33^*;1}&=&-\frac{4}{3} \frac{\alpha_S}{r}~;~~
V_{33^*;8} =  \frac{1}{6} \frac{\alpha_S}{r}~.
\eea
Thus, the most attractive force occurs between a color triplet and an
antitriplet in an overall singlet of color SU(3).  This hierarchy also appears
to hold for the longer-range force, described by a linear potential.  The
slope $k$ of such a potential $V(r) = kr$ between a triplet and an antitriplet
in the color singlet of SU(3) is related to the (approximately universal)
slope $\alpha'$ of Regge trajectories relating masses and spins of mesons
or baryons:
\beq
\alpha(M^2) = \alpha_0 + \alpha' M^2~,~~\alpha' \simeq 0.88~{\rm GeV}^{-2}~,
\eeq
with \cite{Nambu74,StXa} $k = (2 \pi \alpha')^{-1} \simeq 0.18$ GeV$^2$.

A QCD string between a heavy quark $Q$ and antiquark $\bar Q$ has been shown to
break above a separation of about 1.5 fm = 7.5 GeV$^{-1}$ \cite{Rosner:1996xz}
due to production of a light $q \bar q$ pair corresponding to flavor threshold:
\beq
Q \bar Q \to (Q \bar q) + (\bar Q q)~.
\eeq
Let us assume that this string is imparting an impulse in the direction of
the proton or antiproton remnant (i.e., along the beam axis) for the lifetime
of the top quark ($\tau_t=5 \times 10^{-25}$ s).  This is about a tenth of the
time it would take the string to break if its ends separated at the velocity
of light to the critical distance of 1.5 fm:
\beq
\frac{1.5 \times 10^{-15}~{\rm m}}{3 \times 10^8~({\rm m/s})} = 5 \times
10^{-24}~s~.
\eeq

The average impulse imparted to the top quark along the beam axis is then
($c=1$)
\beq
\Delta p_{tz} = k \tau_t = \frac{(0.18~{\rm GeV}^2)(5 \times 10^{-25}~{\rm s})}
{6.582 \times 10^{-25}~{\rm GeV}\cdot{\rm s}} \simeq 0.14~{\rm GeV}~,
\eeq
far too small to cause any noticeable forward-backward asymmetry.  However,
the top quark decays to a bottom quark, whose motion may be affected enough
to cause a noticeable systematic shift in the reconstructed top quark mass, as
we show in the next Section.

\section{EFFECTS ON TOP RECONSTRUCTION}

At the Tevatron, the top quark is produced in a $t \bar t$ pair with a
$m(t \bar t)$ distribution peaked below 400 GeV \cite{CDFAFB,Miet,D0AFB}.
We shall thus neglect its motion with respect to the $t \bar t$ center of mass
in calculating the effect of ``string drag'' on its colored decay product, the
bottom quark.
 
The top quark decays via $t \to b W^+$ with a mean proper lifetime of $5 \times
10^{-25}$ s.  The ensuing $b$ quark carries its color and hence is subject to
the same drag force from the initial proton's remnant.  It is acted upon by
this force for the time it takes the QCD string to break (i.e., to elongate to
its critical length of 1.5 fm), or an average of $5 \times 10^{-24}$ s as noted
above.  The momentum imparted to the $b$ quark is thus about $\Delta p_{bz}
\simeq 1.4$ GeV.  A detailed calculation of the corresponding effect on
reconstructed top quark mass would involve adding $\Delta p_{bz} \simeq 1.4$
GeV to every $b$ quark produced in a Monte Carlo simulation of $t \to b
W^+$.  However, the following simplified estimate captures the essence of the
effect.

Top quark pairs are produced in $\bar p p$ collisions at the Tevatron with a
distribution in rapidity $y$ peaked around $y = 0$ with typical width
$|\Delta y| \simeq 1$.  We shall consider pairs produced with $y=0$ and
shall neglect the motion of each top quark with respect to the pair's center
of mass.  We shall thus consider a top quark at rest decaying to $b W^+$, where
the $b$ is emitted with an angle $\theta^*$ with respect to the beam axis.
With
\beq
m_t = 173~{\rm GeV}~,~~m_b = 4.8~{\rm GeV}~,~~M_W = 80.385~{\rm GeV}~,
\eeq
we find the momenta and energies of the $b$ and $W$ in the top rest frame to be
\beq
|\vec{p}_b| = |\vec{p}_W| = 67.7~{\rm GeV}~,~~E_b = 67.9~{\rm GeV}~,~~
E_W = 105.1~{\rm GeV}~.
\eeq
An impulse $\Delta p_{zb}$ along the proton beam axis causes a change
\beq
\Delta m_t = \frac{|\vec{p}_b|}{E_b} \Delta p_{zb} \cos \theta^*
\eeq
in the reconstructed value of the top quark mass, to lowest order in
$\Delta p_{zb}$.  The shift in the reconstructed top quark mass is thus
very close to the impulse imparted by the string drag, as the velocity
$\beta_b = |\vec{p}_b|/E_b$ of the $b$ quark in the $t$ center of mass is
very nearly 1.

Such an effect, while small, could lead to systematic shifts in reconstructed
top quark masses, depending not only on the angle of emission of the $b$ quark
but on whether the recoiling $W$ decays to two jets or a lepton pair.  We
urge trial analyses in which arbitrary small impulses $\Delta p_{zb}$ are added
to the $b$ quarks in top quark decays, to see if such impulses lead to
consistent top mass reconstruction.  These effects should be less pronounced
at the LHC, where top quark production occurs mainly through $gg \to t \bar t$,
leaving proton remnants of zero triality and hence symmetric interaction
with $t$ and $\bar t$.

\section{EFFECTS ON UNDERLYING EVENT}

The strings in Fig.\ \ref{fig:drag} connecting the $b$ quark with the proton
remnant and the $\bar b$ with the antiproton remnant will undergo further
fragmentation as they stretch, in the manner of a string between a quark and
an antiquark in $e^+ e^- \to q \bar q$ \cite{Artru:1974hr,Andersson:1983ia}.
The products of this fragmentation will have limited transverse momenta with
respect to the string axis.  The underlying event in top pair production thus
may exhibit non-uniformity in the distribution of hadrons in an azimuthal
angle $\phi$ around the beam axis, where $\phi = 0$ is defined by the plane
containing the beam axis and the $b$ quark's direction.  At the Tevatron,
the fragments should be more concentrated in $\phi$ for $b$ quarks making
smaller angles with the proton.

\section{FORWARD BARYON PRODUCTION}

If a heavy quark tends to be produced in the direction of the incident
baryon, as suggested in Fig.\ \ref{fig:drag}, one might expect to see more
$\Lambda_b$ than $\bar \Lambda_b$ in the direction of the proton beam,
and more $\bar \Lambda_b$ than $\Lambda_b$ in the direction of the antiproton
beam, at the Tevatron.  At the LHC, one should
see the ratio of $\bar \Lambda_b$ to $\Lambda_b$ production decrease for
large $|y|$.  The CMS Collaboration finds no significant deviation of this
ratio from a constant in the central region $|y^{\Lambda_b}| < 2$, though the
ratio for $1.5 \le |y^{\Lambda_b}| \le 2$ of $0.67 \pm 0.16 \pm 0.08$ is
consistent with a mild decrease in the largest $|y|$ bin \cite{CMS}.  The
LHCb Collaboration is in an ideal position to extend this measurement to
larger $|y|$, where a string-drag picture would predict a growing predominance
of $\Lambda_b$ over $\bar \Lambda_b$ with increasing $|y|$.

\section{CONCLUSIONS}

The forward-backward asymmetry of top production at the Fermilab Tevatron
remains a mystery in the standard model.  We have examined whether a breakdown
of the factorization assumption could lead the $t$ to preferentially follow the
incident proton, perhaps as a result of unexpectedly large nonperturbative QCD
effects.  In a model in which a QCD string preferentially connects the $t$ to
the proton remnants, it was shown that only a very small momentum is imparted
to the top quark before it decays, and even the momentum imparted to its
daughter $b$ quark by the QCD string is only about 1.4 GeV/$c$ before the
string breaks.  While this is unlikely to account for any notable
forward-backward asymmetry in top production, it could lead to small systematic
shifts in the top quark mass reconstructed by CDF and D0, depending on the
angle between the $b$ quark and the proton beam.  We urge a systematic search
for such effects.

Other symptoms of a ``string-drag'' effects on heavy quark production include
azimuthal asymmetries of the underlying event in top pair production and a
tendency of heavy baryons to be favored over antibaryons in the proton
direction at the Tevatron.

While top pair production is dominated by initial $q \bar q$ states at
the Tevatron, it is mainly governed by gluon-gluon collisions at the LHC, so
the picture of Fig.\ \ref{fig:drag} does not apply there.  Nonetheless, in the
the very forward direction ($|y| > 1.5$) at the LHC, one also expects heavy
baryon production to be favored over antibaryon production.  A systematic
study of how this effect varies from $\Lambda_c$ to $\Lambda_b$ could help
shed light on whether a similar charge asymmetry arises for top production.

Two recent observations relate to the possible enhancement of $A_{FB}$ within
QCD.  (1) Appreciable forward-backward asymmetries in top production can arise
when the leading-order prediction (with no asymmetry) is supplemented with
fragmentation in several Monte Carlo approaches \cite{Skands:2012mm}.  (2)
Setting a renormalization scale using a prescription known as the Principle of
Maximal Conformality \cite{Brodsky:2012ik} improves agreement between
experiment and QCD prediction for $A_{FB}$.

\section*{ACKNOWLEDGMENTS}

We thank S. Stone and L.-T. Wang for helpful discussions.  This work was
supported in part by the United States Department of Energy under Grant No.\
DE-FG02-90ER40560.


\begin{thebibliography}{99}

\bibitem{CDFAFB} T. Aaltonen {\it al.} (CDF Collaboration), Phys.\ Rev.\
D {\bf 83}, 112993 (2011); CDF Conference Notes 10436 and 10584 (2011).

\bibitem{Miet}
D. Mietlicki, Joint Experimental-Theoretical Physics Seminar, Fermilab,
March 30, 2012.

\bibitem{D0AFB} V. M. Abazov {\it et al.} (D0 Collaboration), Phys.\ Rev.\
D {\bf 84}, 112005 (2011).

\bibitem{AFBth} S. Frixione, P. Nason, and G. Ridolfi, JHEP 0709 (2007) 126;
N. Kidonakis, Phys.\ Rev.\ D {\bf 84}, 011504 (2011); V. Ahrens {\it al.},
Phys.\ Rev.\ D {\bf 84}, 074004 (2011); W. Hollik and D. Pagani, Phys.\ Rev.\
D {\bf 84}, 093003 (2011); J. K\"uhn and G. Rodrigo, JHEP 1201 (2012) 063
[arXiv:1109.6830]; A. V. Manohar and M. Trott, arXiv:1201.3926 [hep-ph].

\bibitem{Cao:2010zb} See, e.g., the following and references therein:
Q.~-H.~Cao, D.~McKeen, J.~L.~Rosner, G.~Shaughnessy and C.~E.~M.~Wagner,
  Phys.\ Rev.\ D {\bf 81}, 114004 (2010);
M. Gresham, I. W. Kim, and K. Zurek, arXiv:1102.0018;
D. Duffy, Z. Sullivan, and H. Zhang, arXiv:1203.4489.

\bibitem{ATLAS:2012an} 
  G.~Aad {\it et al.} (ATLAS Collaboration),
  arXiv:1203.4211 [hep-ex].

\bibitem{Chatrchyan:2011hk} 
  S.~Chatrchyan {\it et al.} (CMS Collaboration),
  Phys.\ Lett.\ B {\bf 709}, 28 (2012)
  [arXiv:1112.5100 [hep-ex]].

\bibitem{CMSTOP11030} CMS Collaboration, Public Note CMS PAS TOP-11-030, March
1, 2012.

\bibitem{Kuhn:2011ri} 
  J.~H.~Kuhn and G.~Rodrigo,
  JHEP {\bf 1201}, 063 (2012)
  [arXiv:1109.6830 [hep-ph]].

\bibitem{LissPDG} T. M. Liss and A. Quadt, in K. Nakamura {\it et al.}
(Particle Data Group), J. Phys. G {\bf 37}, 075021 (2010), p.\ 596.

\bibitem{Nambu} Y. Nambu, in {\it Preludes in Theoretical Physics}, edited by
A. De-Shalit, H. Feshbach, and L. Van Hove, North-Holland, Amsterdam, 1966,
p.\ 133.

\bibitem{Lipkin73} H. J. Lipkin, Phys.\ Lett. {\bf 45B}, 267 (1973).

\bibitem{StX} For a simple prescription for evaluating the corresponding
group-theoretic factors see J. L. Rosner in {\it Techniques and Concepts of
High-Energy Physics}, edited by T. Ferbel, NATO Advanced Study Institutes
Series B: Physics, Vol.\ 66 (Plenum, New York, 1981), pp.\ 112-114.

\bibitem{Nambu74} Y. Nambu, Phys.\ Rev.\ D {\bf 10}, 4262 (1974).

\bibitem{StXa} For an elementary discussion see J. L. Rosner \cite{StX}, pp.\
18-20.

\bibitem{Rosner:1996xz} 
  J.~L.~Rosner,
  Phys.\ Lett.\ B {\bf 385}, 293 (1996)
  [hep-ph/9605373].

\bibitem{Artru:1974hr} 
  X.~Artru and G.~Mennessier,
  Nucl.\ Phys.\ B {\bf 70}, 93 (1974).

\bibitem{Andersson:1983ia} 
  B.~Andersson, G.~Gustafson, G.~Ingelman and T.~Sjostrand,
  Phys.\ Rept.\  {\bf 97}, 31 (1983).

\bibitem{CMS}
  CMS Collaboration, arXiv:1205.0594 [hep-ex].

\bibitem{Skands:2012mm} 
  P.~Z.~Skands, B.~R.~Webber and J.~Winter,
  arXiv:1205.1466 [hep-ph].

\bibitem{Brodsky:2012ik} 
  S.~J.~Brodsky and X.~-G.~Wu,
  arXiv:1205.1232 [hep-ph].
\end{thebibliography}
\end{document}